\documentclass[conference,letterpaper]{IEEEtran}
\usepackage[utf8]{inputenc} 
\usepackage[T1]{fontenc}
\usepackage{url}
\usepackage{ifthen}
\usepackage{cite}
\usepackage{tabularx}
\usepackage{booktabs}
\usepackage{caption}
\usepackage{multirow}
\usepackage[cmex10]{amsmath} % Use the [cmex10] option to ensure complicance
\usepackage{mathtools}
\usepackage[shortlabels]{enumitem}
\usepackage{graphicx,epic,eepic,epsfig,latexsym,verbatim,color}
 
\usepackage{amsfonts}       % blackboard math symbols
\usepackage{nicefrac}       % compact symbols for 1/2, etc.
\usepackage{algorithm}
\usepackage{algorithmic}

\usepackage[table]{xcolor}

\usepackage{framed}
\definecolor{shadecolor}{rgb}{0.9,0.90,0.9}
\usepackage{bbm}
 
\usepackage{float}
\usepackage{tikz}
\usetikzlibrary{chains}
\usetikzlibrary{fit}

\usepackage{epsfig}
\usetikzlibrary{shapes.symbols,patterns} % for source symbols
\usepackage{pgfplots}

\usepackage[strict]{changepage}

\usepackage[marginal]{footmisc}
\usepackage{url}

\newtheorem{proposition}{Proposition}

\newtheorem{corollary}[proposition]{Corollary}

\def\squareforqed{\hbox{\rlap{$\sqcap$}$\sqcup$}}
\def\qed{\ifmmode\squareforqed\else{\unskip\nobreak\hfil
\penalty50\hskip1em\null\nobreak\hfil\squareforqed
\parfillskip=0pt\finalhyphendemerits=0\endgraf}\fi}
\def\endenv{\ifmmode\;\else{\unskip\nobreak\hfil
\penalty50\hskip1em\null\nobreak\hfil\;
\parfillskip=0pt\finalhyphendemerits=0\endgraf}\fi}

\newcounter{remark}

\newcounter{example}

\usepackage{graphicx}

\newcommand{\nc}{\newcommand}
\nc{\rnc}{\renewcommand}
\nc{\lbar}[1]{\overline{#1}}
\nc{\bra}[1]{\langle#1|}
\nc{\ket}[1]{|#1\rangle}
\nc{\dketbra}[2]{\vert #1 \rangle \hspace{-.8mm} \rangle \hspace{-.4mm} \langle\hspace{-.8mm}\langle #2 \vert}
\nc{\dbra}[1]{\langle\hspace{-.8mm}\langle #1\vert}
\nc{\dket}[1]{\vert#1\rangle\hspace{-.8mm}\rangle}
\nc{\ketbra}[2]{|#1\rangle\!\langle#2|}
\nc{\braket}[2]{\langle#1|#2\rangle}
\nc{\braandket}[3]{\langle #1|#2|#3\rangle}

\nc{\proj}{{\mathrm{Proj}}}
\nc{\avg}[1]{\langle#1\rangle}
\nc{\rank}{\operatorname{rank}}
\nc{\smfrac}[2]{\mbox{$\frac{#1}{#2}$}}
\nc{\tr}{\operatorname{Tr}}
\nc{\ox}{\otimes}
\nc{\dg}{\dagger}
\nc{\dn}{\downarrow}
\nc{\cA}{{\cal A}}
\nc{\cB}{{\cal B}}
\nc{\cC}{{\cal C}}
\nc{\cD}{{\cal D}}
\nc{\cE}{{\cal E}}
\nc{\cF}{{\cal F}}
\nc{\cG}{{\cal G}}
\nc{\cH}{{\cal H}}
\nc{\cI}{{\cal I}}
\nc{\cJ}{{\cal J}}
\nc{\cK}{{\cal K}}
\nc{\cL}{{\cal L}}
\nc{\cM}{{\cal M}}
\nc{\cN}{{\cal N}}
\nc{\cO}{{\cal O}}
\nc{\cP}{{\cal P}}
\nc{\cQ}{{\cal Q}}
\nc{\cR}{{\cal R}}
\nc{\cS}{{\cal S}}
\nc{\cT}{{\cal T}}
\nc{\cU}{{\cal U}}
\nc{\cV}{{\cal V}}
\nc{\cX}{{\cal X}}
\nc{\cY}{{\cal Y}}
\nc{\cZ}{{\cal Z}}
\nc{\cW}{{\cal W}}

\nc{\rmT}{{\mathrm{T}}}
\nc{\rmD}{{\mathrm{D}}}
\nc{\rmS}{{\mathrm{S}}}
\nc{\rmR}{{\mathrm{R}}}

\nc{\csupp}{{\operatorname{csupp}}}
\nc{\qsupp}{{\operatorname{qsupp}}}
\nc{\var}{{\operatorname{var}}}
\nc{\rar}{\rightarrow}
\nc{\lrar}{\longrightarrow}
\nc{\polylog}{{\operatorname{polylog}}}
\nc{\idop}{{\mathds{1}}}
\nc{\wt}{{\operatorname{wt}}}
\nc{\av}[1]{{\left\langle {#1} \right\rangle}}
\nc{\St}{{\operatorname{St}}}
\nc{\SEP}{{\mathrm{SEP}}}
\nc{\PPT}{{\mathrm{PPT}}}
\nc{\grad}{{\mathrm{grad}}}

\nc{\argmin}{{\operatorname{argmin}}}

\nc{\RR}{{{\mathbb R}}}
\nc{\CC}{{{\mathbb C}}}
\nc{\FF}{{{\mathbb F}}}
\nc{\NN}{{{\mathbb N}}}
\nc{\ZZ}{{{\mathbb Z}}}
\nc{\PP}{{{\mathbb P}}}
\nc{\QQ}{{{\mathbb Q}}}
\nc{\UU}{{{\mathbb U}}}
\nc{\EE}{{{\mathbb E}}}
\nc{\id}{{\operatorname{id}}}
\nc{\LOCC}{{\text{\rm LOCC}}}

\usepackage{thmtools}
\usepackage{thm-restate}
\usepackage{etoolbox}

\begin{document}
\title{Riemannian Optimization for Holevo Capacity}

\author{%
\IEEEauthorblockN{Chengkai Zhu\IEEEauthorrefmark{1}, Renfeng Peng\IEEEauthorrefmark{2}\IEEEauthorrefmark{3}, Bin Gao\IEEEauthorrefmark{2}, Xin Wang\IEEEauthorrefmark{1}}
\IEEEauthorblockA{\IEEEauthorrefmark{1}%
Thrust of Artificial Intelligence, Information Hub, \\The Hong Kong University of Science and Technology (Guangzhou), Guangzhou, China.}
\IEEEauthorblockA{\IEEEauthorrefmark{2}%
State Key Laboratory of Scientific and Engineering Computing, Academy of Mathematics and Systems Science, \\
Chinese Academy of Sciences, Beijing, China.}
\IEEEauthorblockA{\IEEEauthorrefmark{3}%
University of Chinese Academy of Sciences, Beijing, China.\\
Email: gaobin@lsec.cc.ac.cn, felixxinwang@hkust-gz.edu.cn}
}

\maketitle

\begin{abstract}
Computing the classical capacity of a noisy quantum channel is crucial for understanding the limits of communication over quantum channels. However, its evaluation remains challenging due to the difficulty of computing the Holevo capacity and the even greater difficulty of regularization. In this work, we formulate the computation of the Holevo capacity as an optimization problem on a product manifold constructed from probability distributions and their corresponding pure input states for a quantum channel. A Riemannian gradient descent algorithm is proposed to solve the problem, providing lower bounds on the classical capacity of general quantum channels and outperforming existing methods in numerical experiments in both efficiency and scale.
\end{abstract}

\section{Introduction}
Quantum information theory extends classical information theory by incorporating the unique properties of quantum systems, governed by the laws of quantum mechanics. Central to this framework are quantum channels, which describe the evolution of quantum states through noisy communication links. When transmitting classical information through a quantum channel, the classical capacity denotes the maximum rate at which the channel can reliably convey information over an asymptotically large number of channel uses. This concept is essential for understanding quantum communication systems' ultimate limits and for designing protocols that can approach these limits. However, evaluating the classical capacity remains a formidable challenge due to the optimization over high-dimensional quantum state spaces and the need to consider joint measurements on the output states.

The celebrated result by Holevo, Schumacher and Westmoreland~\cite{Holevo1973BoundsFT,Schumacher1997,Holevo1998} showed that the classical capacity of a quantum channel $\cN$ is given by 
\begin{equation}\label{Eq:classical_cap}
    C(\cN) = \lim_{n\rightarrow \infty} \frac{1}{n}\chi(\cN^{\ox n}),
\end{equation}
where
\begin{equation}\label{def:holevo_cap}
    \chi(\cN) \coloneqq \max_{\{p_i, \rho_i\}} H\Big(\sum_{i}p_i \cN(\rho_i)\Big) - \sum_i p_i H\big(\cN(\rho_i)\big)
\end{equation}
is the Holevo capacity of the channel, and $H(\rho)\coloneqq - \tr(\rho\log\rho)$ is the von Neumann entropy of a quantum state~$\rho$. One may recast computing the classical capacity to computing the Holevo capacity if the Holevo capacity is additive, i.e., $\chi(\cN^{\ox n}) = n\chi(\cN)$. For example, $\chi(\cdot)$ is additive for entanglement breaking channels~\cite{Shor_2002}, unital qubit channels~\cite{King_2002}, depolarizing channels~\cite{King2003}, erasure channels~\cite{Bennett_1997}, and Hadamard channels~\cite{king2004,king2006}. However, Hastings~\cite{Hastings_2009} demonstrated that the Holevo capacity is not additive in general, which consequently makes the regularization in Eq.~\eqref{Eq:classical_cap} necessary. As evaluating the Holevo capacity is NP-complete~\cite{beigi2008}, the exact classical capacity of a general quantum channel is computationally intractable although there are several efficiently computable upper bounds~\cite{Wang2016g,Wang2017a,Fang2019a,Leditzky2018}.

Due to the aforementioned challenges, the existing work for computing the Holevo capacity of a general quantum channel is quite sparse. Inspired by the \textit{Blahut--Arimoto algorithm} used for calculating the capacity of classical channels~\cite{Blahut1972,Arimoto1972}, Nagaoka~\cite{Nagaoka1998} proposed a Blahut--Arimoto type algorithm to compute the Holevo capacity of quantum channels. In~\cite{Shor_2003}, Shor adopted a combinatorial optimization technique to approximate the Holevo capacity. There are other numerical approaches (see, e.g.,~\cite{osawa2001numerical,kato2008computational}). However, the theoretical convergence analysis for these methods is unknown.

Notably, Sutter \textit{et al.}~\cite{Sutter_2016} developed a convex optimization method with provable convergence bounds for approximating the Holevo capacity of classical-quantum (cq) channels. 
More recently, quantum versions of the Blahut--Arimoto algorithm were proposed~\cite{Haobo2019,Haobo2019e,Ramakrishnan_2021} for estimating the Holevo capacity of cq channels with convergence guarantees. In addition, Fawzi \textit{et al.}~\cite{Fawzi_2018,Fawzi_2018e} proposed a framework to estimate the Holevo capacity of cq channels via semidefinite programming (SDP), which has favorable runtimes. Despite these advancements on cq channels, algorithms for computing the Holevo capacity of general quantum channels are still lacking.

In this work, we propose a provable Riemannian gradient descent algorithm for lower bounding the Holevo capacity of arbitrary quantum channels. 
We formulate the computation of the Holevo capacity as a non-convex optimization problem defined on a Cartesian product of smooth Riemannian manifolds. 
The proposed framework has various advantages. First, it applies to general quantum channels without any specific restrictions. Second, it reduces the computational complexity of evaluating the Holevo capacity by exploiting the geometric structure of the problem. Third, it scales better for high-dimensional quantum systems compared with existing numerical methods. We demonstrate the efficiency of our approach through numerical simulations on several examples of practical interest, including depolarizing channels, classical-quantum channels, entanglement-breaking channels, Pauli channels, and more general ones. 

%%%%%%%%%%%%%%%%%%%%%%%%%%%%%%%%%%%%%%%%%%%%%%%%%%%%%%%%%%%%
%%%%%%%%%%%%%%%%%%%%%%%%%%%%%%%%%%%%%%%%%%%%%%%%%%%%%%%%%%%%
\section{Quantum states and quantum channels}
A pure quantum state is a $d$-dimensional complex vector $\ket{\psi} \in \mathbb{C}^d$ with $\braket{\psi}{\psi} = 1$. A general state of a quantum system is described by a positive-semidefinite density matrix $\rho \in \mathbb{C}^{d\times d}$ with $\tr \rho = 1$. The standard $n$-simplex is $\Delta^n \coloneqq \{x \in \mathbb{R}^{n+1}: x_0,\dots,x_{n}\geq 0, \sum_{i=0}^n x_i =1\}$ and the relative interior of the simplex is $\Delta_+^{n}\coloneqq \{x\in\RR^{n+1}: x_0,\dots,x_n > 0, \sum_{i=0}^n x_i = 1\}$. The binary entropy is defined by $h_2(p) = -p\log p - (1-p)\log (1-p)$. For two density matrices $\rho$ and $\sigma$, the quantum relative entropy of $\rho$ with respect to $\sigma$ is defined by $D(\rho\|\sigma) = \tr \rho(\log\rho - \log \sigma)$. We denote $[\,\cdot\,]_i$ as the $i$-th element of a vector. Throughout this paper, $\log$ denotes the binary logarithm. A quantum channel $\cN_{A\to B}$ is a linear map from system $A$ to system $B$ that is both completely positive (CP) and trace-preserving (TP). We denote by $\cD(A)$ the set of density matrices on $A$. The dimension of a system $A$ is denoted as $d_A = |A|$. The action of a quantum channel has a Kraus representation, i.e., $\cN(\rho_A) = \sum_k K_k^{} \rho_A K_k^\dagger$ with $\sum_k K_k^\dagger K_k^{} = \mathbb{I}$.

%%%%%%%%%%%%%%%%%%%%%%%%%%%%%%%%%%%%%%%%%%%%%%%%%%%%%%%%%%%%
%%%%%%%%%%%%%%%%%%%%%%%%%%%%%%%%%%%%%%%%%%%%%%%%%%%%%%%%%%%%
\section{Riemannian optimization for Holevo capacity}
In this section, we formulate the problem of computing the Holevo capacity as an optimization problem on a smooth manifold and propose a Riemannian gradient descent algorithm. Riemannian optimization aims to solve the optimization problems on smooth manifolds by exploiting the Riemannian geometry of manifolds and preserving the manifold structure; see~\cite{absil2009optimization,boumal2023intromanifolds} for an overview.

For a quantum channel $\cN_{A\rightarrow B}$, the maximum in Eq.~\eqref{def:holevo_cap} can be achieved by pure-state ensembles of cardinality at most~$d_A^2$, i.e., by $\{p_i, \ket{\psi_i}\}_{i=0}^{d_A^2-1}$~\cite{Schumacher1997}. We observe that $\ket{\psi_i}$ lies in a complex unit sphere $\rmS^{d_A-1}\coloneqq \{\ket{\psi} \in \CC^{d_A}:\braket{\psi}{\psi} = 1\}$, which is an \textit{embedded submanifold} of~$\CC^{d_A}$. Additionally, the probability $\{p_i\}_{i=0}^{d_A^2-1}$ lies in a relative interior of the simplex, i.e., $\Delta_{+}^{d_A^2-1}$, which is an embedded submanifold of~$\RR^{d_A^2}$. Consequently, computing the Holevo capacity of $\cN_{A\rightarrow B}$ can be formulated by the following Riemannian optimization problem
\begin{equation}\label{Eq:holevo_cap_pure}
   \chi(\cN) = -\min_{m\in \cM_{d_A^2+1}} f_{\cN}(m),
\end{equation}
where the search space
\begin{figure}[htbp]
    \centering
    \includegraphics[width=.92\linewidth]{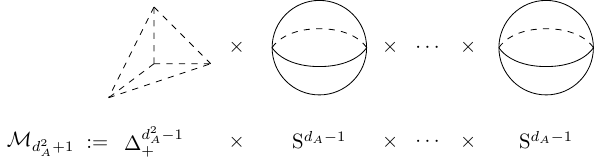}
\end{figure}\\
is a Cartesian product of $(d_A^2+1)$ manifolds, and $m = (p, \ket{\psi_0}, \dots,\ket{\psi_{d_A^2-1}})$ is a point in $\cM_{d_A^2+1}$. The \textit{Holevo cost function} is defined by
\begin{equation*}
    f_{\cN}(m) \coloneqq \sum_{i=0}^{d_A^2-1} p_i H\left(\cN(\ketbra{\psi_i}{\psi_i})\right) - H\Big( \cN\Big(\sum_{i=0}^{d_A^2-1} p_i \ketbra{\psi_i}{\psi_i}\Big)\Big).
\end{equation*}

\begin{figure}[htbp]
    \centering
    \includegraphics[width=0.75\linewidth]{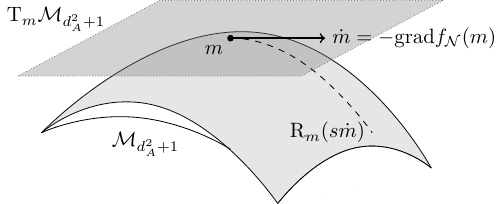}
    \caption{Illustration of optimization on the manifold $\cM_{d_A^2+1}$.}
    \label{fig:optim_manifold}
\end{figure}

In order to facilitate Riemannian optimization algorithms on the product manifold $\cM_{d_A^2+1}$, which involves choosing a search direction, a stepsize, and projection onto the manifold, we delve into the Riemannian geometry of~$\cM_{d_A^2+1}$; see~\cite{gao2023optimization} and Fig.~\ref{fig:optim_manifold} for an illustration.

Given $m\in\cM_{d_A^2+1}$, the tangent space $\rmT_{m}\cM_{d_A^2+1}$ is given by 
\[\rmT_{p}\Delta_+^{d_A^2-1}\times\rmT_{\ket{\psi_0}}\rmS^{d_A-1}\times\cdots\times\rmT_{\ket{\psi_{d_A^2-1}}}\rmS^{d_A-1},
\]
where $\rmT_{p} \Delta_{+}^{d_A^2-1} = \{\dot{p}\in\RR^{d_A^2}:\sum_{i=0}^{d_A^2-1} \dot{p}_i = 0\}$ and $\rmT_{\ket{\psi}}\rmS^{d_A-1} = \{\ket{\dot{\psi}}\in\CC^{d_A}: \braket{\psi}{\dot{\psi}} = 0\}$ are tangent spaces of $\Delta_{+}^{d_A^2-1}$ and $\rmS^{d_A-1}$, respectively. Given tangent vectors $\dot{m}_0=(\dot{p},\ket{\dot{\psi}_0},\dots,\ket{\dot{\psi}_{d_A^2-1}}),~\dot{m}_1=(\dot{q},\ket{\dot{\phi}_0},\dots,\ket{\dot{\phi}_{d_A^2-1}})\in\rmT_{m}\cM_{d_A^2+1}$, the Riemannian metric $\langle\cdot,\cdot\rangle_{m}$ is defined by $\langle \dot{m}_0, \dot{m}_1\rangle_{m}\coloneqq \sum_{i=0}^{d_A^2-1}(\dot{p}_i\dot{q}_i+\braket{\dot{\psi}_i}{\dot{\phi}_i})$. 

The \textit{Riemannian gradient} $\grad f(m)$ of a smooth function $f: \cM_{d_A^2+1}\to\mathbb{R}$ is the unique tangent vector at $m$ such that $\rmD f(m)[\dot{m}] = \langle \dot{m},{\grad f(m)}\rangle_m$ holds for all $\dot{m} \in \rmT_{m}\cM_{d_A^2+1}$, where $\rmD f(m)$ is the differential of $f$ at~$m$. The Riemannian gradient can be computed by projections,
\begin{equation}\label{Eq:Riemannian_gradient}
\begin{aligned}
    \grad f(m) =& \Big(\proj_{p}(\partial_{p}\Bar{f}(m)),~\proj_{\ket{\psi_0}}\big(\partial_{\ket{\psi_0}} \Bar{f}(m)\big),\\
    &\qquad\quad \dots, \proj_{\ket{\psi_{d_A^2-1}}}\big(\partial_{\ket{\psi_{d_A^2-1}}} \Bar{f}(m)\big)\Big),
\end{aligned}
\end{equation}
where $[\proj_{p}(x)]_i = x_i-\sum_{i=0}^{d_A^2-1}x_i/d_A^2$ for $x\in\mathbb{R}^{d_A^2}$, $\proj_{\ket{\psi}}(\ket{v}) = (\mathbb{I}-\ketbra{\psi}{\psi})\ket{v}$ for $\ket{v}\in\mathbb{C}^{d_A}$, and $\Bar{f}:\RR^{d_A^2}\times\CC^{d_A}\times\cdots\times\CC^{d_A}\rightarrow \RR$ and $f$ coincide on $\cM_{d_A^2+1}$. Note that the Riemannian gradient of $f$ on $\cM_{d_A^2+1}$ can be alternatively computed by the Riemannian gradients on each component of the product manifold. 

A \emph{retraction} map $\rmR_{m}:\rmT_{m} \cM_{d_A^2+1}\to\cM_{d_A^2+1}$, aiming to ``project'' a tangent vector onto the manifold, is given by
\begin{equation}\label{Eq:retraction}
    \rmR_{m}(\dot{m}) = \Big(\rmR_{p}(\dot{p}), \rmR_{\ket{\dot{\psi}_0}}(\ket{\dot{\psi}_0}), \dots , \rmR_{\ket{\dot{\psi}_{d_A^2-1}}}(\ket{\dot{\psi}_{d_A^2-1}})\Big)
\end{equation}
for a tangent vector $\dot{m}\in\rmT_{m}\cM_{d_A^2+1}$, where $\rmR_{p}$ and $\rmR_{\ket{\psi}}(\ket{\dot{\psi}}) = (\ket{\psi} + \ket{\dot{\psi}})/\|\ket{\psi} + \ket{\dot{\psi}}\|$ are retractions on $\Delta_{+}^{d_A^2-1}$ and $\rmS^{d_A-1}$, respectively. Note that the retraction $\rmR_{p}(\dot{p})$ can be computed by $\rmR_{p}(\dot{p})=\hat{p}/\sum_{i=0}^{d_A^2-1}\hat{p}_i$ with $\hat{p}_i=p_i+\dot{p}_i+\dot{p}_i^2/(2p_i)$.

Specifically, we compute the Riemannian gradient of the Holevo cost function $f_{\cN}(m)$ on the product manifold $\cM_{d_A^2+1}$ as follows.

\begin{proposition}\label{prop:qq_de}
For a quantum channel $\cN_{A\rightarrow B}$ and a pure state ensemble $\{p_i, \ket{\psi_i}\}_{i=0}^{d_A^2-1}$, denote $\sigma_i = \cN(\ketbra{\psi_i}{\psi_i})$ and $\sigma = \sum_{i} p_i \sigma_i $. The Riemannian gradient of the Holevo cost function at $m = (p, \ket{\psi_0}, \dots,\ket{\psi_{d_A^2-1}})$ is computed by 
\begin{equation*}
\begin{aligned}
    [\grad f_{\cN}(m)]_0 \!&= q,\\
    \big[\grad f_{\cN}(m)\big]_i \!&= 2p_i \Big[\cN^{\dagger} \big(\log \sigma - \log \sigma_i \big) + D(\sigma_i\|\sigma) \mathbb{I}\Big]\ket{\psi_i}
\end{aligned}
\end{equation*}
for $i=1,2,\dots,d_A^2$, where $q_j = 1-D(\sigma_j \| \sigma) + p_j(\sum_{k=0}^{d_A^2-1} p_k D(\sigma_k \| \sigma) - 1)$ for $j=0,\dots, d_A^2-1$.
\end{proposition}
\vspace{2mm}
\begin{IEEEproof}
In order to compute the derivatives, we introduce the following argument
\begin{equation}\label{Eq:dir_de}
    \frac{\partial}{\partial \ket{\ell}} \bra{\psi}\mathsf{W}(\ket{\psi})\ket{\psi} =\bra{\psi}\frac{\partial}{\partial \ket{\ell}} \mathsf{W}(\ket{\psi}) \ket{\psi} + 2\bra{\ell} \mathsf{W}(\ket{\psi})\ket{\psi},
\end{equation}
that computes the directional derivative of the function $\bra{\psi}\mathsf{W}(\ket{\psi})\ket{\psi}$ along a vector $\ket{\ell}\in\mathbb{C}^{n}$, where $\mathsf{W}:\mathbb{C}^{n}\to\mathbb{C}^{n\times n}$ is a mapping. 
Given a quantum channel $\cN_{A\rightarrow B}$ with Kraus operators $\{K_i\}$, we denote the Holevo cost function as $f(\{p_i, \ket{\psi_i}\}) = f_{\cN}^{(1)}(\{p_i, \ket{\psi_i}\}) - f_{\cN}^{(2)}(\{p_i, \ket{\psi_i}\})$ where
\begin{equation*}
\begin{aligned}
    &f_{\cN}^{(1)} = \sum_{i_1} p_{i_1} \sum_{k_1} \bra{\psi_{i_1}} K_{k_1}^\dagger \big(\log \sigma\big)K_{k_1}\ket{\psi_{i_1}},\\
    &f_{\cN}^{(2)} =\sum_i p_i \sum_{k_1} \bra{\psi_i}K_{k_1}^\dagger \big(\log \sigma_i \big) K_{k_1}\ket{\psi_i},
\end{aligned}
\end{equation*}
and $\sigma\coloneqq \sum_{k,i} p_{i} K_{k}\ketbra{\psi_{i}}{\psi_{i}} K_{k}^\dagger,~\sigma_i\coloneqq \sum_{k} K_{k}\ketbra{\psi_{i}}{\psi_{i}} K_{k}^\dagger$. In the following, we compute the partial derivatives of $f_{\cN}^{(1)}$ and $f_{\cN}^{(2)}$ with respect to the states $\ket{\psi_i}$, respectively. 

For $f_{\cN}^{(1)}$, according to Eq.~\eqref{Eq:dir_de}, we can calculate the directional derivative along the state $\ket{\ell_i}$
\begin{equation*}
\begin{aligned}
    \langle\partial_{\ket{\psi_i}} f_{\cN}^{(1)}\ket{\ell_i} &= 2p_i \bra{\ell_i} \sum_{k_1} K_{k_1}^\dagger \left(\log \sigma \right) K_{k_1}\ket{\psi_i}\\
    +&\sum_{i_1}p_{i_1}\bra{\psi_{i_1}} \frac{\partial}{\partial \ket{\ell_i}}\sum_{k_1} K_{k_1}^\dagger \left(\log \sigma \right) K_{k_1}\ket{\psi_{i_1}}.
\end{aligned}
\end{equation*}
Notice that
\begin{equation*}
\begin{aligned}
    &\bra{\psi_{i_1}} \frac{\partial}{\partial \ket{\ell_i}}\sum_{k_1} K_{k_1}^\dagger \Big(\log \sum_{k_2, i_2} p_{i_2} K_{k_2}\ketbra{\psi_{i_2}}{\psi_{i_2}} K_{k_2}^\dagger\Big) K_{k_1}\ket{\psi_{i_1}}\\
    &= \tr \Big[\Big( \sum_{k_2} p_i K_{k_2}(\ketbra{\ell_{i}}{\psi_{i}} + \ketbra{\psi_{i}}{\ell_{i}}) K_{k_2}^\dagger\Big)\sigma^{-1} \sigma_{i_1}\Big]\\
    &= \tr\sum_{k_2} \Big[p_i (\ketbra{\ell_{i}}{\psi_{i}}+\ketbra{\psi_{i}}{\ell_{i}}) K_{k_2}^\dagger \sigma^{-1} \sigma_{i_1} K_{k_2}\Big]\\
    &= p_i\sum_{k_2} \Big\langle \ketbra{\ell_{i}}{\psi_{i}} + \ketbra{\psi_{i}}{\ell_{i}}, ~Z^{(1)}_{i_1}\Big\rangle\\
    &= p_i\sum_{k_2} \bra{\ell_i} (Z^{(1)}_{i_1}+(Z^{(1)}_{i_1})^\dagger) \ket{\psi_{i}},
\end{aligned}
\end{equation*}
where we define $Z^{(1)}_{i_1} \coloneqq K_{k_2}^\dagger \sigma_{i_1} \sigma^{-1} K_{k_2}$. 
Therefore, we have
\begin{equation*}
    \partial_{\ket{\psi_i}} f_{\cN}^{(1)} = 2p_i \cN^\dagger \left(\log \sigma\right)\ket{\psi_i} + \sum_{i_1,k_2} p_i p_{i_1}(Z^{(1)}_{i_1} + (Z^{(1)}_{i_1})^\dagger)\ket{\psi_{i}}.
\end{equation*}
Since $\sum_{i_1,k_2} p_{i_1}Z^{(1)}_{i_1} = \sum_{k_2} K_{k_2}^\dagger \sigma \sigma^{-1} K_{k_2} = \mathbb{I}$, it follows that
\begin{equation}\label{Eq:f1_deri}
    \partial_{\ket{\psi_i}} f_{\cN}^{(1)} = 2p_i \big(\cN^\dagger \left(\log \sigma\right) +\mathbb{I}\big) \ket{\psi_{i}}.
\end{equation}
Followed by a similar calculation, we yield the directional derivative of $f_{\cN}^{(2)}$ as 
$\langle\partial_{\ket{\psi_i}} f_{\cN}^{(2)}\ket{\ell_i} = 2p_i \bra{\ell_i} \cN^\dagger \left(\log \sigma_i \right) \ket{\psi_i} + p_i\bra{\psi_{i}} \frac{\partial}{\partial\ket{\ell_i}} \cN^\dagger \left(\log \sigma_i \right) \ket{\psi_{i}}$,
and thus
\begin{equation*}
    \partial_{\ket{\psi_i}} f_{\cN}^{(2)} = 2p_i \cN^\dagger \left(\log \sigma_i \right)\ket{\psi_i} + \sum_{k_2} p_i(Z^{(2)}_i + (Z^{(2)}_i)^\dagger)\ket{\psi_i},
\end{equation*}
where we define $Z^{(2)}_i\coloneqq K_{k_2}^\dagger \sigma_i \sigma_i^{-1} K_{k_2}$. 
Since $\sum_{k_2} Z^{(2)}_i = \sum_{k_2} K_{k_2}^\dagger \sigma_i \sigma_i^{-1} K_{k_2} = \mathbb{I}$, it follows that 
\begin{equation}\label{Eq:f2_deri}
    \partial_{\ket{\psi_i}} f_{\cN}^{(2)} = 2p_i \big(\cN^\dagger \left(\log \sigma_i\right) +\mathbb{I}\big) \ket{\psi_{i}}.
\end{equation}
Combining Eq.~\eqref{Eq:f1_deri} and Eq.~\eqref{Eq:f2_deri}, we have
\begin{equation*}
    \partial_{\ket{\psi_i}} f_{\cN} = 2p_i \cN^\dagger \big(\log \sigma - \log \sigma_i \big) \ket{\psi_i}.
\end{equation*}

In the following, we compute the partial derivatives of $f_{\cN}^{(1)}$ and $f_{\cN}^{(2)}$ with respect to $p_i$, respectively.
\begin{equation*}
\begin{aligned}
    \partial_{p_i} f_{\cN}^{(1)} &= \tr \sigma_i\big(\log \sigma \big) + \sum_{i_1}p_{i_1}\frac{\partial}{\partial p_i}\tr \sigma_i \big(\log \sigma\big)\\
    &= \tr \sigma_i\big(\log \sigma \big) + \sum_{i_1}p_{i_1} \tr \sigma_i \sigma^{-1} \sigma_i\\
    &= \tr \sigma_i\big(\log \sigma \big) + \tr \sigma\sigma^{-1} \sigma_i\\
    &= \tr \sigma_i\big(\log \sigma \big) + 1.
\end{aligned}
\end{equation*}
Further, we have $\partial_{p_i} f_{\cN}^{(2)} = \tr \sigma_i\big(\log \sigma_i \big)$. Thus, we have $\partial_{p_i} f_{\cN} = 1-D(\sigma_i\|\sigma)$. We note that this partial derivative has also been calculated for cq channels in Ref.~\cite{Kerry2024}. 
Recall the Riemannian gradient~\eqref{Eq:Riemannian_gradient}. It holds that 
\begin{equation*}
\begin{aligned}
    [\grad f_{\cN}(m)]_i &= 2p_i \Big[\cN^\dagger \big(\log \sigma - \log \sigma_i \big) - \\
    &\quad \sum_{k_1} \bra{\psi_i} K_{k_1}^\dagger \big(\log \sigma - \log \sigma_i \big) K_{k_1}\ket{\psi_i}\Big]\ket{\psi_i}\\
    &= 2p_i \Big[\cN^{\dagger} \big(\log \sigma - \log \sigma_i \big) + D(\sigma_i\|\sigma)\Big] \ket{\psi_i}.
\end{aligned}
\end{equation*}
Similarly, we have $[\grad f_{\cN}(m)]_0=q$ with $q_i = 1-D(\sigma_i \| \sigma) + p_i(\sum_j p_j D(\sigma_j \| \sigma) - 1)$.
\end{IEEEproof}
\vspace{2mm}

Particularly, Proposition~\ref{prop:qq_de} can be applied to a classical-quantum channel (cq channel) $\Phi: \cX \rightarrow \cD(B), x\mapsto \rho_x$, where $\cX$ is a given input alphabet, $B$ is a quantum system, and $\{\rho_x\}_{x\in\cX}$ is a given set of quantum states.

\vspace{2mm}
\begin{corollary}
For a classical-quantum channel $\Phi:x\mapsto \rho_x$ and $\rho = \sum_x p_x \rho_x$, the Riemannian gradient of the Holevo cost function $f_{\Phi}$ at $p$ is given by 
\begin{equation*}
    \big[\grad f_{\Phi}(p)\big]_x \!=\! 1-D(\rho_x \| \rho) + p_x\Big(\sum_j p_j D(\rho_x \| \rho) - 1\Big).
\end{equation*}
\end{corollary}

By using the Riemannian gradient $\grad f_{\cN}(m)$, we propose a Riemannian gradient descent (RGD) algorithm to solve the
optimization problem~\eqref{Eq:holevo_cap_pure} in Algorithm~\ref{alg: RGD}, where the stepsize is determined by the Armijo backtracking line search. Since the problem is typically nonconvex, Algorithm~\ref{alg: RGD} is not guaranteed to find the global optimal solution. Nevertheless, it is an efficient method to compute points in $\cM_{d_A^2+1}$ that satisfies necessary optimality conditions, and thus, gives a lower bound on $\chi(\cdot)$ which is of significant practical interest. A point $m^*\in\cM_{d_A^2+1}$ is a first-order \textit{critical point} of~\eqref{Eq:holevo_cap_pure} if $\grad f_{\cN}(m^*)=0$. The RGD algorithm computes a point $m^{(t)}$ such that $\|\grad f_{\cN}(m^{(t)})\| \leq \epsilon$ within $t = \cO(1/\epsilon^2)$ steps~\cite{Boumal_2018}.

\begin{algorithm}[htbp]
\caption{Riemannian gradient descent algorithm (RGD)}\label{alg: RGD}
\begin{algorithmic}[1]
      \REQUIRE Initial guess $m^{(0)}\in\cM_{d_A^2+1},t=0$. 
      \WHILE{the stopping criteria are not satisfied}
          \STATE Compute $\dot{m}^{(t)} = -\grad f_{\cN}(m^{(t)})$ by~\eqref{Eq:Riemannian_gradient}.
          \STATE Compute a stepsize $s^{(t)}$.
          \STATE Update $m^{(t+1)}= \rmR_{m^{(t)}} (s^{(t)}\dot{m}^{(t)})$ by~\eqref{Eq:retraction}; $t= t+1$.
      \ENDWHILE
      \ENSURE $m^{(t)}\in\cM_{d_A+1}$.
\end{algorithmic}
\end{algorithm}

It is worth noting that Algorithm~\ref{alg: RGD} can be employed for general quantum channels without any specific restrictions. We remark that the Holevo cost function $f_{\cN}$ is not smooth in general as $\cN(\ketbra{\psi_i}{\psi_i})$ can be singular. Nevertheless, in practice, we compute the Riemannian gradient of $f$ under the channel $\cN' = (1-\delta)\cN + \delta \cD$ to avoid the singularity where $\cD$ is a fully depolarizing channel and $\delta\in(0,1)$ is a sufficiently small parameter, e.g., $\delta = 10^{-9}$. 

Recall that the Holevo capacity of a tensor product of quantum channels is essential to the classical capacity. In the following, we explore the Riemannian gradient property for the case of tensor product of quantum channels on product input states of which the proof is similar to that in Proposition~\ref{prop:qq_de}.
% Appendix~\ref{app:qq_grad}.
\begin{proposition}\label{prop:2copy_qq_de}
For quantum channels $\cN_{A\rightarrow B}, \cN'_{A'\rightarrow B'}$ and respective pure state ensemble $\{p_i, \ket{\psi_i}\}, \{q_j, \ket{\phi_j}\}$, the Riemannian gradient of the Holevo cost function $f_{\cN\ox \cN'}$ with respect to $\ket{\psi_i}\ox \ket{\phi_j}$ is given by 
\begin{equation*}
\begin{aligned}
    \grad f_{\cN\ox\cN'}(\ket{\psi_i}\ox \ket{\phi_j}) &= \grad f_{\cN}(\ket{\psi_i}) \ox (1-q_j)\ket{\psi_j}\\
    &~~~+ (1-p_i) \ket{\psi_i} \ox \grad f_{\cN'}(\ket{\phi_j}).
\end{aligned}
\end{equation*}
\end{proposition}
We observe from Proposition~\ref{prop:2copy_qq_de} that if $\ket{\hat{\psi_i}}$ and $\ket{\hat{\phi_j}}$ satisfy $\|\grad f_{\cN}(\ket{\hat{\psi_i}})\|\leq \epsilon$ and $\|\grad f_{\cN'}(\ket{\hat{\phi_j}})\|\leq \epsilon$ respectively, it follows that $\|\grad f_{\cN\ox\cN'}(\ket{\hat{\psi_i}}\ox \ket{\hat{\phi_j}})\|\leq 2\epsilon$. This implies that, to explore the superadditivity of the Holevo capacity, a product input state is not a desirable initial point for optimization as it could trap the algorithm in local minima.

%%%%%%%%%%%%%%%%%%%%%%%%%%%%%%%%%%%%%%%%%%%%%%%%%%%%%%%%%%%%
%%%%%%%%%%%%%%%%%%%%%%%%%%%%%%%%%%%%%%%%%%%%%%%%%%%%%%%%%%%%
\section{Numerical results}
In this section, we present numerical results on the performance of Algorithm~\ref{alg: RGD} for approximating the Holevo capacities of different channels. The implementation of Riemannian algorithms is based on a Matlab toolbox Manopt~\cite{boumal2014manopt}. The proposed method is terminated if $\sqrt{\langle\grad f(m),\grad f(m)\rangle_m}\leq 10^{-6}$. We compare the proposed RGD algorithm with the SDP-based method~\cite{Fawzi_2018e} and the first-order method in Ref.~\cite{Sutter_2016}. All the simulations are performed on a 2.4 GHz AMD EPYC 7532 32-core Processor with 256 GB RAM under Matlab R2024a. The implementation codes are available at~\cite{coderepo}.

%%%%%%%%%%%%%%%%%%%%%%%%%%%%%%%%%%%%%%%%%%%%%%%%%%%%%%%%%%%%
\paragraph{Depolarizing channel}
The $d$-dimensional depolarizing channel with a noise parameter $\lambda$ is given by $\cD_{\lambda}^{(d)}(\rho) = (1-\lambda)\rho + \lambda \mathbb{I}_d/d$. King~\cite{King2003} showed that the Holevo capacity of $\cD_{\lambda}^{(d)}$ is given by 
\begin{equation}\label{Eq:holevocap_depo}
    \chi(\cD_{\lambda}^{(d)}) = \log d + \lambda' \log \lambda' + \frac{\lambda(d-1)}{d} \log \left(\frac{\lambda}{d}\right),
\end{equation}
where $\lambda' = 1-\lambda+\lambda/d$. Table~\ref{tab:depo} shows the absolute error between the analytical value in Eq.~\eqref{Eq:holevocap_depo} and the value obtained by the RGD algorithm for different $d$ with $\lambda=1/3$. Observe that the proposed method demonstrates high accuracy in estimating the Holevo capacity. Compared to the method in~\cite{Sutter_2016}, for the case of $d=2$, their approach required a time on the order of $10^4$ seconds to achieve accuracy on the order of $10^{-6}$ relative to the true value. In contrast, the proposed RGD algorithm obtains an accuracy on the order of $10^{-14}$ within less than a second of computation time.

\begin{table}[ht]
\centering
\captionof{table}{Classical capacity of $d$-dimensional depolarizing channels with noise parameter $\lambda =1/3$.}
\begin{tabular}{rrrrrr}
\toprule % \textbf{Comp. time (s)}\textbf{Abs. err.}
$d$ & \multicolumn{1}{c}{time (s)} & \multicolumn{1}{c}{abs. err.} & $d$ & \multicolumn{1}{c}{time (s)} & \multicolumn{1}{c}{abs. err.}\\
\cmidrule(r){1-3} \cmidrule(r){4-6}
2 & 0.04 & 4.32e-14 & 3 & 0.15 & 2.05e-12\\
8 & 0.61 & 1.48e-13 & 7 & 0.91 & 1.95e-12\\
16 & 10.22 & 1.46e-12 & 21 & 42.32 & 1.31e-11\\
\bottomrule
\end{tabular}
\label{tab:depo}
\end{table}

%%%%%%%%%%%%%%%%%%%%%%%%%%%%%%%%%%%%%%%%%%%%%%%%%%%%%%%%%%%%
\paragraph{Classical-quantum channel}

Ref.~\cite{Fawzi_2018e} proposed a method for estimating the Holevo capacity of cq channels via SDP. We compare computation time for the SDP method~\cite{Fawzi_2018e} and our RGD algorithm for the Holevo capacity cq channels in Table~\ref{tab:cq}. Our method improves efficiency and can scale up for evaluating large dimensional cq channels. Notably, for cq channels with a binary input alphabet $\cX = \{0, 1\}$ and pure output states $\Phi(0)$ and $\Phi(1)$, the Holevo capacity is analytically given by $h_2((1+\epsilon)/2)$~\cite{Holevo_2012}. Our method achieves an absolute error of the order of $10^{-16}$ in estimating this capacity, significantly outperforming the SDP method, which yields an error of $10^{-6}$. Further, we generate some random pure states $\{\rho_{x}\}$ with different sizes of $\cX$ and compare the computation time of the two methods in Table~\ref{tab:cq}. The proposed RGD algorithm demonstrates superior computational efficiency and scalability for channels with large input and output dimensions compared to the SDP method~\cite{Fawzi_2018e}.

\begin{table}[ht]
\centering
\captionof{table}{Holevo capacity of random classical-quantum channels with different input sizes and output dimensions.}
\begin{tabular}{rrrrrr}
\toprule
\multirow{2}{*}{$|\cX|$} & \multirow{2}{*}{$d_B$} &  \multicolumn{2}{c}{\centering \textbf{SDP method~\cite{Fawzi_2018e}}} & \multicolumn{2}{c}{\centering\textbf{RGD}} \\
\cmidrule(r){3-4}\cmidrule(r){5-6}
 & & \multicolumn{1}{c}{time (s)} & \multicolumn{1}{c}{est. value} & \multicolumn{1}{c}{time (s)} & \multicolumn{1}{c}{est. value} \\ 
\midrule
2 & 20 & 17.74 & 0.99462457 & 0.03 & 0.99462458 \\
2 & 100 & - & - & 0.07 & 0.99964658 \\
10 & 20 & 40.01 & 2.96086525 & 0.19 & 2.96086706 \\
10 & 100 & - & - & 0.48 & 3.26025080 \\
100 & 100 & - & - & 4.99 & 5.92762710 \\
100 & 500 & - & - & 138.45 & 6.49899737 \\
\bottomrule
\end{tabular}
\label{tab:cq}
\end{table}

%%%%%%%%%%%%%%%%%%%%%%%%%%%%%%%%%%%%%%%%%%%%%%%%%%%%%%%%%%%%
\paragraph{Entanglement breaking channel}
A quantum channel $\cN_{A\rightarrow B}$ is called an \textit{entanglement-breaking} channel~\cite{Horodecki_2003}, if $\id_R \ox \cN_{A\rightarrow B}(\rho_{RA})$ is separable for all auxiliary quantum systems $R$ and input states $\rho_{RA}$ where $\id_R$ is the identity channel. A channel $\cN$ is entanglement-breaking if and only if it has a Kraus representation $\{K_i\}$ with each $K_i$ being rank one~\cite{Horodecki_2003}, i.e., $K_i = \ketbra{w_i}{v_i}$. It is shown that for entanglement-breaking channel $\cN$, the Holevo capacity is additive~\cite{Shor_2002}, and thus $C(\cN) = \chi(\cN)$. 

We construct $d$-dimensional entanglement-breaking channels with $d$ Kraus operators by generating $d$-dimensional Haar random pure states $\{\ket{w_i}\}_{i=0}^{d-1}$ and a Haar random orthonormal basis $\{\ket{v_i}\}_{i=0}^{d-1}$. Then we compute their classical capacity using the RGD algorithm. Note that the classical capacity of an entanglement-breaking channel is equal to its entanglement-assisted classical capacity which can be computed by the SDP-based method in Ref.~\cite{Fawzi_2018e}. We compute the estimation value and computation time of the SDP method~\cite{Fawzi_2018e} and the RGD algorithm, shown in Table~\ref{tab:eb}. We observe that the RGD algorithm provides more precise estimations when its output exceeds that of the SDP method, as it consistently yields a lower bound on the Holevo capacity. Furthermore, it scales better for high-dimensional general quantum channels compared to the SDP method, which rapidly encounters memory limitations.

\begin{table}[ht]
\centering
\captionof{table}{Classical capacity of $d$-dimensional random entanglement-breaking channels.}
\begin{tabular}{rrrrr}
\toprule
\multirow{2}{*}{$d_A$} &  \multicolumn{2}{c}{\centering \textbf{SDP method~\cite{Fawzi_2018e}}} & \multicolumn{2}{c}{\centering\textbf{RGD}} \\
\cmidrule(r){2-3}\cmidrule(r){4-5}
 & \multicolumn{1}{c}{time (s)} & \multicolumn{1}{c}{est. value} & \multicolumn{1}{c}{time (s)} & \multicolumn{1}{c}{est. value} \\ 
\midrule
2	& 4.71 	& 0.66841354 & 0.32  & 0.66841365 \\
3	& 1293.79 	& 1.15927364 & 2.93  & 1.15939173 \\
6	& -	& - & 21.68  & 1.89803233 \\
10	& -	& - & 426.55  & 2.78430087 \\
\bottomrule
\end{tabular}
\label{tab:eb}
\end{table}

%%%%%%%%%%%%%%%%%%%%%%%%%%%%%%%%%%%%%%%%%%%%%%%%%%%%%%%%%%%%
\paragraph{Pauli channels} 
We denote single-qubit Pauli matrices as 
$X=\begin{pmatrix}
\begin{smallmatrix}
    0 & 1 \\
    1 & 0
\end{smallmatrix}
\end{pmatrix},
Y=\begin{pmatrix}
\begin{smallmatrix}
0 & -i \\
i & 0
\end{smallmatrix}
\end{pmatrix},
Z =\begin{pmatrix}
\begin{smallmatrix}
1 & 0 \\
0 & -1
\end{smallmatrix}
\end{pmatrix}$.
The qubit Pauli channel is described by 
\begin{equation*}
\cP_{p}(\rho) = (1-q)\rho + p_X X\rho X^{\dagger} + p_Y Y\rho Y^{\dagger} + p_Z Z\rho Z^{\dagger},    
\end{equation*}
where $q=p_X+p_Y+p_Z$. To compare our method with the approach in Ref.~\cite{Sutter_2016}, we estimate the Holevo capacity of a Pauli channel with $p_X = 1/7, p_Y=1/10$ and $p_Z= 1/4$. Numerical experiments demonstrate that the RGD algorithm achieves an estimated value of $0.20024$ in $0.19$ seconds, with a relative gap of order $10^{-5}$ from the upper bound provided in~\cite{Sutter_2016}. In contrast, their method requires the order of $10^5$ seconds to obtain a posteriori error (the gap between lower and upper bounds) of order $10^{-3}$. This example also demonstrates the computational efficiency of the proposed method for general quantum channels.

\begin{figure}[htbp]
    \centering
    \includegraphics[width=0.9\linewidth]{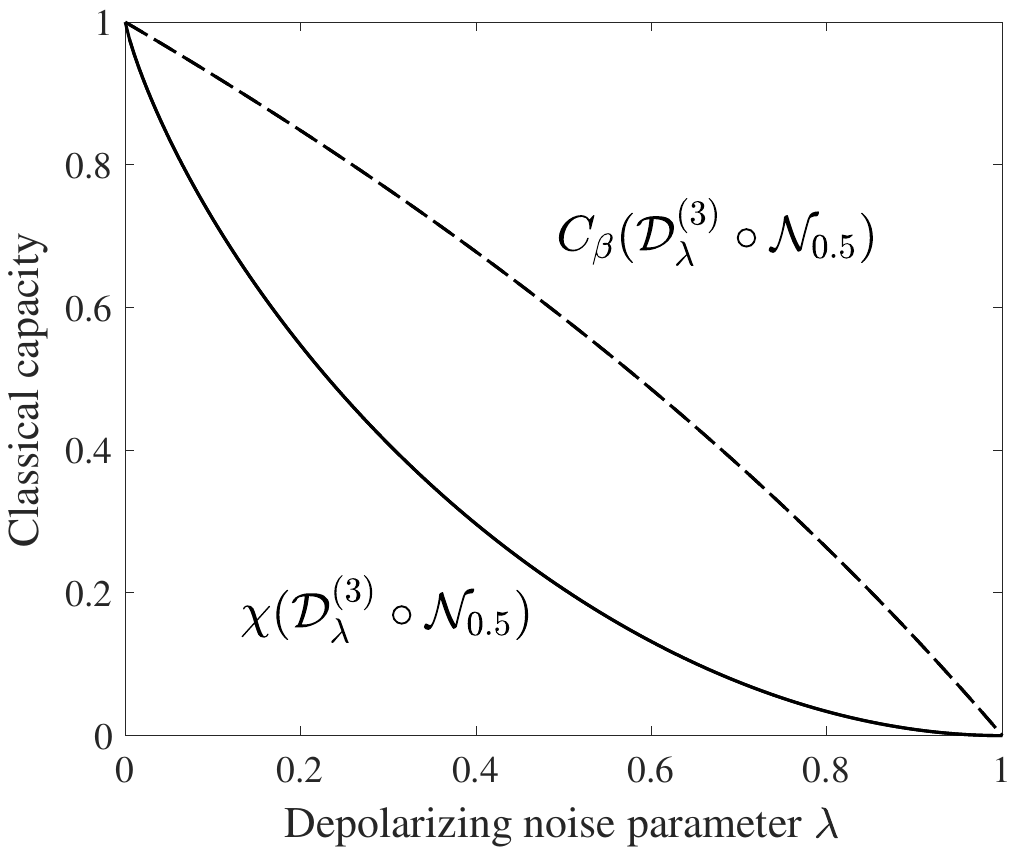}
    \caption{Upper and lower bounds on the classical capacity of the quantum channel $\cD_\lambda^{(3)}\circ\cN_{0.5}$ with varying $\lambda$.}
    \label{fig:WDchannel}
\end{figure}

%%%%%%%%%%%%%%%%%%%%%%%%%%%%%%%%%%%%%%%%%%%%%%%%%%%%%%%%%%%%
\paragraph{Special qutrit channel}
Wang and Duan~\cite{Wang_2018} introduced a class of qutrit quantum channels $\cN_{\alpha}(\rho) = E_{\alpha}\rho E_{\alpha}^\dagger + D_{\alpha}\rho D_{\alpha}^\dagger, 0<\alpha \leq \pi/4$ with $E_{\alpha} = \sin \alpha \ketbra{0}{1} + \ketbra{1}{2}, D_{\alpha} = \cos\alpha \ketbra{2}{1} + \ketbra{1}{0}$, for which the quantum Lov\'asz number is strictly larger than the entanglement assisted zero-error capacity. Also, this class of channel was applied to explore the platypus of the quantum channel zoo~\cite{Leditzky2023}. The Holevo capacity of this class of channels is shown to be 1~\cite{Wang2016g}. Hence, to explore a more intrinsic scenario, we consider the composition of $\cN_{\alpha}$ with a 3-dimensional depolarizing channel $\cD_\lambda^{(3)}$, for which the Holevo capacity is not trivially to know. We estimate its classical capacity by lower bounding the Holevo capacity and compared it with the strong converse bound $C_{\beta}$~\cite{Wang2016g} on the classical capacity. Fig.~\ref{fig:WDchannel} illustrates the lower bound on $\chi(\cD_\lambda^{(3)}\circ\cN_{0.5})$ via the RGD algorithm and $C_{\beta}(\cD_\lambda^{(3)}\circ\cN_{0.5})$ computed via SDP~\cite{Wang2016g} when $\lambda$ varies.

We can see that the RGD algorithm provides an effective lower bound on the Holevo capacity of general quantum channels. Together with the SDP upper bounds, we have more tools to explore the Holevo capacity and its regularization, which can advance our understanding of the classical capacity of general channels.

%%%%%%%%%%%%%%%%%%%%%%%%%%%%%%%%%%%%%%%%%%%%%%%%%%%%%%%%%%%%
%%%%%%%%%%%%%%%%%%%%%%%%%%%%%%%%%%%%%%%%%%%%%%%%%%%%%%%%%%%%
\section{Conclusion}
We have developed a Riemannian gradient descent algorithm for computing the Holevo capacity of a general quantum channel by reformulating it as an optimization problem on the Cartesian product of complex unit spheres and the probability simplex. Numerical simulations demonstrate that our method achieves high precision while offering superior computational efficiency and scalability compared to existing approaches. The proposed method scales effectively for large-dimensional quantum channels, e.g., depolarizing channels with $d=20$, cq channels with $|\cX| = 100$ and $d_B = 500$. We hope it could provide a way for identifying explicit examples of quantum channels that violate the additivity of the Holevo capacity, i.e., $\chi(\cN_0 \ox \cN_1) > \chi(\cN_0) + \chi(\cN_1)$, for which only existence proofs are currently known~\cite{Hastings_2009}. We anticipate that this work will stimulate further development of Riemannian optimization methods tailored to the Holevo capacity and shed light on more optimization problems~\cite{XWthesis,Tavakoli2024,zhu2024unified,casanova2024} in quantum information.

\section*{Acknowledgment}
Chengkai Zhu and Renfeng Peng contributed equally to this work. This work was partially supported by the National Key R\&D Program of China (Grant No.~2024YFE0102500), the National Natural Science Foundation of China (Grant. No.~12447107), and the Guangdong Provincial Quantum Science Strategic Initiative (Grant No.~GDZX2403008, GDZX2403001).

%%%%%%
\bibliographystyle{IEEEtran}
\bibliography{arxiv_v1}

\begin{thebibliography}{10}
\providecommand{\url}[1]{#1}
\csname url@samestyle\endcsname
\providecommand{\newblock}{\relax}
\providecommand{\bibinfo}[2]{#2}
\providecommand{\BIBentrySTDinterwordspacing}{\spaceskip=0pt\relax}
\providecommand{\BIBentryALTinterwordstretchfactor}{4}
\providecommand{\BIBentryALTinterwordspacing}{\spaceskip=\fontdimen2\font plus
\BIBentryALTinterwordstretchfactor\fontdimen3\font minus \fontdimen4\font\relax}
\providecommand{\BIBforeignlanguage}[2]{{%
\expandafter\ifx\csname l@#1\endcsname\relax
\typeout{** WARNING: IEEEtran.bst: No hyphenation pattern has been}%
\typeout{** loaded for the language `#1'. Using the pattern for}%
\typeout{** the default language instead.}%
\else
\language=\csname l@#1\endcsname
\fi
#2}}
\providecommand{\BIBdecl}{\relax}
\BIBdecl

\bibitem{Holevo1973BoundsFT}
\BIBentryALTinterwordspacing
A.~S. Holevo, ``Bounds for the quantity of information transmitted by a quantum communication channel,'' 1973. [Online]. Available: \url{https://api.semanticscholar.org/CorpusID:118312737}
\BIBentrySTDinterwordspacing

\bibitem{Schumacher1997}
\BIBentryALTinterwordspacing
B.~Schumacher and M.~D. Westmoreland, ``Sending classical information via noisy quantum channels,'' \emph{Phys. Rev. A}, vol.~56, pp. 131--138, Jul 1997. [Online]. Available: \url{https://link.aps.org/doi/10.1103/PhysRevA.56.131}
\BIBentrySTDinterwordspacing

\bibitem{Holevo1998}
A.~Holevo, ``The capacity of the quantum channel with general signal states,'' \emph{IEEE Transactions on Information Theory}, vol.~44, no.~1, pp. 269--273, 1998.

\bibitem{Shor_2002}
\BIBentryALTinterwordspacing
P.~W. Shor, ``Additivity of the classical capacity of entanglement-breaking quantum channels,'' \emph{Journal of Mathematical Physics}, vol.~43, no.~9, p. 4334–4340, Sep. 2002. [Online]. Available: \url{http://dx.doi.org/10.1063/1.1498000}
\BIBentrySTDinterwordspacing

\bibitem{King_2002}
\BIBentryALTinterwordspacing
C.~King, ``Additivity for unital qubit channels,'' \emph{Journal of Mathematical Physics}, vol.~43, no.~10, p. 4641–4653, Oct. 2002. [Online]. Available: \url{http://dx.doi.org/10.1063/1.1500791}
\BIBentrySTDinterwordspacing

\bibitem{King2003}
------, ``The capacity of the quantum depolarizing channel,'' \emph{IEEE Transactions on Information Theory}, vol.~49, no.~1, pp. 221--229, 2003.

\bibitem{Bennett_1997}
\BIBentryALTinterwordspacing
C.~H. Bennett, D.~P. DiVincenzo, and J.~A. Smolin, ``Capacities of quantum erasure channels,'' \emph{Physical Review Letters}, vol.~78, no.~16, p. 3217–3220, Apr. 1997. [Online]. Available: \url{http://dx.doi.org/10.1103/PhysRevLett.78.3217}
\BIBentrySTDinterwordspacing

\bibitem{king2004}
\BIBentryALTinterwordspacing
C.~King, ``An application of a matrix inequality in quantum information theory,'' 2004. [Online]. Available: \url{https://arxiv.org/abs/quant-ph/0412046}
\BIBentrySTDinterwordspacing

\bibitem{king2006}
\BIBentryALTinterwordspacing
C.~King, K.~Matsumoto, M.~Nathanson, and M.~B. Ruskai, ``Properties of conjugate channels with applications to additivity and multiplicativity,'' 2006. [Online]. Available: \url{https://arxiv.org/abs/quant-ph/0509126}
\BIBentrySTDinterwordspacing

\bibitem{Hastings_2009}
\BIBentryALTinterwordspacing
M.~B. Hastings, ``Superadditivity of communication capacity using entangled inputs,'' \emph{Nature Physics}, vol.~5, no.~4, p. 255–257, Mar. 2009. [Online]. Available: \url{http://dx.doi.org/10.1038/nphys1224}
\BIBentrySTDinterwordspacing

\bibitem{beigi2008}
\BIBentryALTinterwordspacing
S.~Beigi and P.~W. Shor, ``On the complexity of computing zero-error and {Holevo} capacity of quantum channels,'' 2008. [Online]. Available: \url{https://arxiv.org/abs/0709.2090}
\BIBentrySTDinterwordspacing

\bibitem{Wang2016g}
\BIBentryALTinterwordspacing
X.~Wang, W.~Xie, and R.~Duan, ``Semidefinite programming strong converse bounds for classical capacity,'' \emph{IEEE Transactions on Information Theory}, vol.~64, no.~1, pp. 640--653, jan 2018. [Online]. Available: \url{http://ieeexplore.ieee.org/document/8012535/}
\BIBentrySTDinterwordspacing

\bibitem{Wang2017a}
\BIBentryALTinterwordspacing
X.~Wang, K.~Fang, and M.~Tomamichel, ``On converse bounds for classical communication over quantum channels,'' \emph{IEEE Transactions on Information Theory}, vol.~65, no.~7, pp. 4609--4619, jul 2019. [Online]. Available: \url{http://arxiv.org/abs/1709.05258}
\BIBentrySTDinterwordspacing

\bibitem{Fang2019a}
K.~Fang and H.~Fawzi, ``Geometric {R{\'{e}}nyi} divergence and its applications in quantum channel capacities,'' \emph{Communications in Mathematical Physics}, vol. 384, no.~3, pp. 1615--1677, jun 2021.

\bibitem{Leditzky2018}
\BIBentryALTinterwordspacing
F.~Leditzky, E.~Kaur, N.~Datta, and M.~M. Wilde, ``Approaches for approximate additivity of the {Holevo} information of quantum channels,'' \emph{Physical Review A}, vol.~97, no.~1, p. 012332, jan 2018. [Online]. Available: \url{https://link.aps.org/doi/10.1103/PhysRevA.97.012332}
\BIBentrySTDinterwordspacing

\bibitem{Blahut1972}
R.~Blahut, ``Computation of channel capacity and rate-distortion functions,'' \emph{IEEE Transactions on Information Theory}, vol.~18, no.~4, pp. 460--473, 1972.

\bibitem{Arimoto1972}
S.~Arimoto, ``An algorithm for computing the capacity of arbitrary discrete memoryless channels,'' \emph{IEEE Transactions on Information Theory}, vol.~18, no.~1, pp. 14--20, 1972.

\bibitem{Nagaoka1998}
H.~Nagaoka, ``Algorithms of {Arimoto--Blahut} type for computing quantum channel capacity,'' in \emph{Proceedings. 1998 IEEE International Symposium on Information Theory (Cat. No.98CH36252)}, 1998, pp. 354--.

\bibitem{Shor_2003}
\BIBentryALTinterwordspacing
P.~W. Shor, ``Capacities of quantum channels and how to find them,'' \emph{Mathematical Programming}, vol.~97, no.~1, p. 311–335, Jul. 2003. [Online]. Available: \url{http://dx.doi.org/10.1007/s10107-003-0446-y}
\BIBentrySTDinterwordspacing

\bibitem{osawa2001numerical}
S.~Osawa and H.~Nagaoka, ``Numerical experiments on the capacity of quantum channel with entangled input states,'' \emph{IEICE transactions on fundamentals of electronics, communications and computer sciences}, vol.~84, no.~10, pp. 2583--2590, 2001.

\bibitem{kato2008computational}
K.~Kato, M.~Oto, H.~Imai, and K.~Imai, ``Computational geometry analysis of quantum state space and its applications,'' \emph{Generalized Voronoi Diagram: A Geometry-Based Approach to Computational Intelligence}, pp. 67--108, 2008.

\bibitem{Sutter_2016}
\BIBentryALTinterwordspacing
D.~Sutter, T.~Sutter, P.~Mohajerin~Esfahani, and R.~Renner, ``Efficient approximation of quantum channel capacities,'' \emph{IEEE Transactions on Information Theory}, vol.~62, no.~1, p. 578–598, Jan. 2016. [Online]. Available: \url{http://dx.doi.org/10.1109/TIT.2015.2503755}
\BIBentrySTDinterwordspacing

\bibitem{Haobo2019}
H.~Li and N.~Cai, ``A {Blahut--Arimoto} type algorithm for computing classical-quantum channel capacity,'' in \emph{2019 IEEE International Symposium on Information Theory (ISIT)}, 2019, pp. 255--259.

\bibitem{Haobo2019e}
\BIBentryALTinterwordspacing
------, ``Computing the classical-quantum channel capacity: experiments on a {Blahut--Arimoto} type algorithm and an approximate solution for the binary inputs, two-dimensional outputs channel,'' 2019. [Online]. Available: \url{https://arxiv.org/abs/1905.08235}
\BIBentrySTDinterwordspacing

\bibitem{Ramakrishnan_2021}
\BIBentryALTinterwordspacing
N.~Ramakrishnan, R.~Iten, V.~B. Scholz, and M.~Berta, ``Computing quantum channel capacities,'' \emph{IEEE Transactions on Information Theory}, vol.~67, no.~2, p. 946–960, Feb. 2021. [Online]. Available: \url{http://dx.doi.org/10.1109/TIT.2020.3034471}
\BIBentrySTDinterwordspacing

\bibitem{Fawzi_2018}
\BIBentryALTinterwordspacing
H.~Fawzi, J.~Saunderson, and P.~A. Parrilo, ``Semidefinite approximations of the matrix logarithm,'' \emph{Foundations of Computational Mathematics}, vol.~19, no.~2, p. 259–296, Mar. 2018. [Online]. Available: \url{http://dx.doi.org/10.1007/s10208-018-9385-0}
\BIBentrySTDinterwordspacing

\bibitem{Fawzi_2018e}
\BIBentryALTinterwordspacing
H.~Fawzi and O.~Fawzi, ``Efficient optimization of the quantum relative entropy,'' \emph{Journal of Physics A: Mathematical and Theoretical}, vol.~51, no.~15, p. 154003, Mar. 2018. [Online]. Available: \url{http://dx.doi.org/10.1088/1751-8121/aab285}
\BIBentrySTDinterwordspacing

\bibitem{absil2009optimization}
P.-A. Absil, R.~Mahony, and R.~Sepulchre, in \emph{Optimization Algorithms on Matrix Manifolds}.\hskip 1em plus 0.5em minus 0.4em\relax Princeton University Press, 2009.

\bibitem{boumal2023intromanifolds}
\BIBentryALTinterwordspacing
N.~Boumal, \emph{An introduction to optimization on smooth manifolds}.\hskip 1em plus 0.5em minus 0.4em\relax Cambridge University Press, 2023. [Online]. Available: \url{https://www.nicolasboumal.net/book}
\BIBentrySTDinterwordspacing

\bibitem{gao2023optimization}
B.~Gao, R.~Peng, and Y.-x. Yuan, ``Optimization on product manifolds under a preconditioned metric,'' \emph{arXiv preprint arXiv:2306.08873}, 2023.

\bibitem{Kerry2024}
K.~He, J.~Saunderson, and H.~Fawzi, ``A bregman proximal perspective on classical and quantum {Blahut--Arimoto} algorithms,'' \emph{IEEE Transactions on Information Theory}, vol.~70, no.~8, pp. 5710--5730, 2024.

\bibitem{Boumal_2018}
\BIBentryALTinterwordspacing
N.~Boumal, P.-A. Absil, and C.~Cartis, ``Global rates of convergence for nonconvex optimization on manifolds,'' \emph{IMA Journal of Numerical Analysis}, vol.~39, no.~1, p. 1–33, Feb. 2018. [Online]. Available: \url{http://dx.doi.org/10.1093/imanum/drx080}
\BIBentrySTDinterwordspacing

\bibitem{boumal2014manopt}
\BIBentryALTinterwordspacing
N.~Boumal, B.~Mishra, P.-A. Absil, and R.~Sepulchre, ``Manopt, a {Matlab} toolbox for optimization on manifolds,'' \emph{The Journal of Machine Learning Research}, vol.~15, no.~1, pp. 1455--1459, 2014. [Online]. Available: \url{http://jmlr.org/papers/v15/boumal14a.html}
\BIBentrySTDinterwordspacing

\bibitem{coderepo}
The code used in this work is available at {\url{https://github.com/Chengkai-Zhu/Riemannian-Opt-for-Holevo-Capacity}} and is currently under polishing. Updates will be provided soon.

\bibitem{Holevo_2012}
\BIBentryALTinterwordspacing
A.~S. Holevo and V.~Giovannetti, ``Quantum channels and their entropic characteristics,'' \emph{Reports on Progress in Physics}, vol.~75, no.~4, p. 046001, Mar. 2012. [Online]. Available: \url{http://dx.doi.org/10.1088/0034-4885/75/4/046001}
\BIBentrySTDinterwordspacing

\bibitem{Horodecki_2003}
\BIBentryALTinterwordspacing
M.~Horodecki, P.~W. Shor, and M.~B. Ruskai, ``Entanglement breaking channels,'' \emph{Reviews in Mathematical Physics}, vol.~15, no.~06, p. 629–641, Aug. 2003. [Online]. Available: \url{http://dx.doi.org/10.1142/S0129055X03001709}
\BIBentrySTDinterwordspacing

\bibitem{Wang_2018}
\BIBentryALTinterwordspacing
X.~Wang and R.~Duan, ``Separation between quantum {Lovász} number and entanglement-assisted zero-error classical capacity,'' \emph{IEEE Transactions on Information Theory}, vol.~64, no.~3, p. 1454–1460, Mar. 2018. [Online]. Available: \url{http://dx.doi.org/10.1109/TIT.2018.2794391}
\BIBentrySTDinterwordspacing

\bibitem{Leditzky2023}
F.~Leditzky, D.~Leung, V.~Siddhu, G.~Smith, and J.~A. Smolin, ``The platypus of the quantum channel zoo,'' \emph{IEEE Transactions on Information Theory}, vol.~69, no.~6, pp. 3825--3849, 2023.

\bibitem{XWthesis}
\BIBentryALTinterwordspacing
X.~Wang, ``Semidefinite optimization for quantum information,'' \emph{PhD thesis}, 2018. [Online]. Available: \url{https://opus.lib.uts.edu.au/handle/10453/127996}
\BIBentrySTDinterwordspacing

\bibitem{Tavakoli2024}
\BIBentryALTinterwordspacing
A.~Tavakoli, A.~Pozas-Kerstjens, P.~Brown, and M.~Ara{\'{u}}jo, ``Semidefinite programming relaxations for quantum correlations,'' \emph{Reviews of Modern Physics}, vol.~96, no.~4, p. 045006, dec 2024. [Online]. Available: \url{https://link.aps.org/doi/10.1103/RevModPhys.96.045006}
\BIBentrySTDinterwordspacing

\bibitem{zhu2024unified}
\BIBentryALTinterwordspacing
X.~Zhu, C.~Zhang, Z.~An, and B.~Zeng, ``Unified framework for calculating convex roof resource measures,'' 2024. [Online]. Available: \url{https://arxiv.org/abs/2406.19683}
\BIBentrySTDinterwordspacing

\bibitem{casanova2024}
\BIBentryALTinterwordspacing
M.~Casanova, K.~Ohki, and F.~Ticozzi, ``Finding quantum codes via {R}iemannian optimization,'' 2024. [Online]. Available: \url{https://arxiv.org/abs/2407.08423}
\BIBentrySTDinterwordspacing

\end{thebibliography}

\end{document}